\documentstyle[amssymb,12pt,thmsa,sw20rui]{article}
\input{tcilatex}

\begin{document}

\author{Hongya Liu\thanks{%
Email: hyliu@dlut.edu.cn}$\smallskip $ \ and \ Qiuyang Zhang \\
%EndAName
Department of Physics, Dalian University of Technology, \\
Dalian, 116024, P.R. China\smallskip}
\title{Time-Dependent Solution for a Star Immersed in a Background Radiation}
\date{}
\maketitle

\begin{abstract}
We study a time-dependent and spherically-symmetric solution with a
star-like source. We show that this solution can be interpreted as an
exterior solution of a contracting star which has a decreasing temperature
and is immersed in a homogenous and isotropic background radiation.
Distribution of the temperature in the fields and close-to Schwarzschild
approximation of the solution are studied. By identifying the radiation with
the cosmic background one, we find that the close-to-Schwarzschild
approximate solution is valid in a wide range in our solar system. Possible
experimental tests of the solution are discussed briefly.

PACS number(s): 04.20.-q, 04.50.+h
\end{abstract}

\section{INTRODUCTION}

It is well known that spherically-symmetric sources such as stars can be
modelled in the simplest way by the interior and exterior Schwarzschild
solutions. But to include the radiation outside a star, more complicated
solutions are required.\cite{MacCbook} These include the Vaidya metric which
uses a retarded time coordinate to describe a radiating atmosphere,\cite
{Vaidya53} the metrics of Herrera and coworkers wherein spheres of matter
are matched to exterior spacetimes,\cite{Herrera84}$^{\text{-}}$\cite
{Herrera92} and the metrics of Glass and Krisch which extend the Vaidya
solution to include both a radiation field and a string fluid.\cite{Glass98}
Recently, Liu and Wesson presented a new kind of solutions in which the
metric is time-dependent (but not of the Vaidya form) and the
energy-momentum tensor is of the form of a perfect fluid for radiation plus
a radial heat flow.\cite{LiuJMP01} Clearly, this solution describes sources
which are time-dependent and spherically symmetric. However, we wish to know
specifically what kind of sources the solution represents. In this paper we
will show that it can describe {\bf exterior} fields of a contracting (or
expanding) star immersed in a homogenous and isotropic background radiation.

\section{4D SOLUTION DERIVED FROM 5D SOLUTIONS}

Campbell's theorem says that any 4D Einstein solution with a source can be
locally embedded in a 5D manifold {\it without} sources whose field
equations in terms of the Ricci tensor are $R_{AB}=0$.\cite{Campbell} (Here
and elsewhere lower case Greek letter run 0,123 and uppercase Latin letters
run 0, 123, 4, and we use units $c=1$.) A major application of Campbell's
theorem is to study embeddings of known 4D Einstein solutions in 5D Ricci
flat manifolds.\cite{Ponce88} Another application of the theorem is to
generate new 4D solutions from known 5D Ricci flat solutions.\cite{LiuJMP01}$%
^{,}$\cite{Liu92} In the following we will show briefly how a new 4D
solution is generated in Ref.[7]. For the purpose of convenience, some of
the equations and notations in Ref. [7] will be re-expressed.

In Reference [7], Liu and Wesson presented a class of 5D solutions. This
class of solutions is time-dependent in 5D and spherically symmetric in 3D
with the 5D metric being 
\begin{equation}
dS^2=B(r)dt^2-\left( 1-\lambda t\right) ^2\left[ A(r)dr^2+r^2\left( d\theta
^2+\sin ^2\theta d\varphi ^2\right) \right] -\left( 1-\lambda t\right)
^{-4}dy^2\;,  \label{5metr}
\end{equation}
where $\lambda $ is a constant and the two functions $B$ and $A$ are
determined by 
\begin{eqnarray}
\frac{B^{\prime }}B &=&3\lambda ^2r\frac AB+\frac{A-1}r\;,  \label{Bprime} \\
\frac{A^{\prime }}A &=&3\lambda ^2r\frac AB-\frac{A-1}r\;,  \label{Aprime}
\end{eqnarray}
where a prime denotes derivative with respect to $r$. Clearly, (\ref{Bprime}%
) and (\ref{Aprime}) can be solved, at least numerically, by imposing
boundary conditions on $B(r)$ and $A(r)$. This 5D solution satisfies the 5D
equations $R_{AB}=0$ and therefore is 5D empty. However, the 4D part of the
5D metric (\ref{5metr}), together with the two equations (\ref{Bprime}) and (%
\ref{Aprime}), defines a 4D solution as shown in the following:\cite
{LiuJMP01} 
\begin{equation}
ds^2=B(r)dt^2-\left( 1-\lambda t\right) ^2\left[ A(r)dr^2+r^2\left( d\theta
^2+\sin ^2\theta d\varphi ^2\right) \right] \quad ,  \label{4metr}
\end{equation}
\begin{eqnarray}
\frac{B^{\prime }}B &=&3\lambda ^2r\frac AB+\frac{A-1}r\;,  \label{Bprime2}
\\
\frac{A^{\prime }}A &=&3\lambda ^2r\frac AB-\frac{A-1}r\;,  \label{Aprime2}
\end{eqnarray}
\begin{equation}
8\pi GT_{\alpha \beta }\equiv G_{\alpha \beta }\;,  \label{Einstein}
\end{equation}
where $T_{\alpha \beta }$ is an effective or induced energy-momentum tensor
with 
\begin{eqnarray}
8\pi GT_0^0 &=&\frac{6\lambda ^2}{B\left( 1-\lambda t\right) ^2}\;,
\label{T00} \\
8\pi GT_1^1 &=&8\pi GT_2^2=8\pi GT_3^3=-\frac{2\lambda ^2}{B\left( 1-\lambda
t\right) ^2}\;,  \label{T11} \\
8\pi GT_0^1 &=&\frac{\lambda B^{\prime }}{AB\left( 1-\lambda t\right) ^3}\;.
\label{T01}
\end{eqnarray}
It was also shown in Ref.[7] that this $T_{\alpha \beta }$ can be modeled as
a perfect fluid plus a radial heat flow, 
\begin{equation}
T_{\alpha \beta }=(\rho +p)u_\alpha u_\beta -pg_{\alpha \beta }+q_\alpha
u_\beta +u_\alpha q_\beta \quad ,  \label{T-heat}
\end{equation}
were $\rho $ is the mass density, $p$ is the pressure, $u_\alpha =\left(
u^0,0,0,0\right) $ is the 4-velocity, $q_\alpha =(0,q^1,0,0)$ is the
heat-flux vector, and $u_\alpha $ and $q_\alpha $ obey the orthogonality
condition $q_\alpha u^\alpha =0$. Then equations (\ref{T00})-(\ref{T01})
yield 
\begin{eqnarray}
\rho  &=&3p=\frac{3\lambda ^2}{4\pi GB\left( 1-\lambda t\right) ^2}\;,
\label{rho} \\
q^1 &=&\frac{\lambda B^{\prime }}{8\pi GAB^{3/2}(1-\lambda t)^3}\quad .
\label{heat}
\end{eqnarray}
Equations (\ref{4metr})-(\ref{heat}) constitute a complete set of the 4D
solution, from which we see that the equation of state of the 4D fluid is $%
\rho =3p$, so it represents a radiation or extra-relativistic particles
accompanied by a radial heat flow.

\section{HEAT FLOW AND TEMPERATURE OF THE FIELDS}

The induced 4D energy-momentum tensor (\ref{T-heat}) describes a
thermodynamical system in which the radial heat current $q_\alpha $ implies
a radial temperature gradient. Now we wish to calculate this temperature
distribution over the fields. The generalized relativistic relation between
heat current and temperature gradient can be found in Refs.[11,12] with 
\begin{equation}
q^\alpha =-\kappa \left( T,_\mu -\stackrel{\cdot }{u}_\mu T\right) h^{\alpha
\mu }\;,\quad \stackrel{\cdot }{u}_\mu \equiv u_{\mu ;\nu }u^\nu \;,
\label{q-T1}
\end{equation}
where $\kappa $ is the coefficient of the thermal conductivity and $%
h^{\alpha \mu }$ is the projection tensor, 
\begin{equation}
h^{\alpha \beta }=u^\alpha u^\beta -g^{\alpha \beta }\;.  \label{project}
\end{equation}
To calculate equations (\ref{q-T1}), we calculate $\stackrel{\cdot }{u}_\mu $
firstly. With use of (\ref{4metr}), we find the only non-vanishing $%
\stackrel{\cdot }{u}_\mu $ being $\stackrel{\cdot }{u}_1=-B^{\prime }/(2B)$.
Furthermore, we assume $T=T(t,r)$. Then equations (\ref{q-T1}) reduce to $%
q^0=q^2=q^3=0$ and 
\begin{equation}
q^1=-\frac \kappa {A\left( 1-\lambda t\right) ^2}\left( T,_1+\frac{B^{\prime
}}{2B}T\right) \;.  \label{q-T2}
\end{equation}
Combining (\ref{heat}) and (\ref{q-T2}), we get 
\begin{equation}
-\kappa \left( T,_1+\frac{B^{\prime }}{2B}T\right) =\frac{\lambda B^{\prime }%
}{8\pi G(1-\lambda t)B^{3/2}}\;.  \label{T-eq}
\end{equation}
We find that this equation can be integrated, giving an exact solution being 
\begin{equation}
\sqrt{B(r)}T(t,r)=\sqrt{B_R}T_R(t)+\frac \lambda {8\pi G\kappa (1-\lambda
t)}\ln \frac{B_R}{B(r)}\;,  \label{T-solu}
\end{equation}
where $R$ is a constant radius, and $B_R$ and $T_R(t)$ are values of $B(r)$
and $T(t,r)$ at $r=R$, respectively. From (\ref{heat}) we see that $\lambda
=0$ corresponds to thermal equilibrium $q^\alpha =0$. Then, from (\ref
{T-solu}), if $\lambda =0$ we get $\sqrt{B}T=$constant. Thus we recover the
conclusion that thermal equilibrium corresponds not to constant temperature,
but to the redshifted temperature distribution $\sqrt{g_{00}}T=$constant.%
\cite{MTWbook} Generally we have $\lambda \neq 0$ and the exact solution (%
\ref{T-solu}) determines both the space distribution and the time variation
of the temperature $T$ over the fields.

\section{CLOSE-TO-SCHWARZSCHILD APPROXIMATION}

The close-to-Schwarzschild approximation of the solution (\ref{4metr}) was
given in Ref.[7]. We find that it can be re-expressed in the following form: 
\begin{eqnarray}
ds^2 &=&B(r)dt^2-\left( 1-\lambda t\right) ^2\left[ A(r)dr^2+r^2\left(
d\theta ^2+\sin ^2\theta d\varphi ^2\right) \right] \quad ,  \label{clos-to1}
\\
B &=&1-\frac{2GM}r+2\lambda ^2r^2+O\left( \varepsilon ^3\right) \quad ,
\label{clos-to2} \\
A^{-1} &=&1-\frac{2GM}r-\lambda ^2r^2+O\left( \varepsilon ^3\right) \quad \;.
\label{clos-to3}
\end{eqnarray}
Here $\varepsilon $ is a small quantity of the order of the Newtonian
potential $GM/r$, $O\left( \varepsilon ^3\right) $ are terms of the order of 
$\varepsilon ^3$ or higher, and we have assumed, for practical usage, that $%
\lambda r$ is also a small quantity of the order $\varepsilon $, i.e., 
\begin{equation}
\left| \lambda \right| r\sim GM/r\sim \varepsilon \ll 1\quad .  \label{small}
\end{equation}
To verify the accuracy of this solution, one can calculate $B^{\prime }/B$
and $A^{\prime }/A$ firstly, and then substitute them into the two equations
(\ref{Bprime}) and (\ref{Aprime}). Thus the approximate solutions (\ref
{clos-to1})-(\ref{clos-to3}) are correct in the range (\ref{small}), or,
equivalently, in 
\begin{equation}
GM\ll r\ll \left| \lambda \right| ^{-1}\quad .  \label{r-range}
\end{equation}
We also conclude that (\ref{clos-to1})-(\ref{clos-to3}) are accurate up to
the second order of $\varepsilon $ and give back to the Schwarzschild
solution if $\lambda =0$. So generally we can interpret the solution as an
exterior solution of a star-like source. We will show, in the next section,
that in a wide range in our solar system the condition (\ref{small}), or,
equivalently, (\ref{r-range}), is satisfied.

Consider now the equation (\ref{rho}), which, by (\ref{clos-to2}), reduces
to 
\begin{equation}
\rho =3p=\frac{3\lambda ^2}{4\pi G\left( 1-\lambda t\right) ^2}\left[ 1+%
\frac{2GM}r+O\left( \varepsilon ^2\right) \right] \quad .  \label{rho-appr}
\end{equation}
Here $\rho $ and $p$ constitute a perfect fluid with the equation of state $%
\rho =3p$, implying a property for radiation or ultra-relativistic
particles. From (\ref{rho-appr}) we see that neglecting higher-order terms
in the square bracket in the RHS of (\ref{rho-appr}), the densities $\rho $
and $p$ are homogenous and isotropic. Thus we find that one can not
interpret the fluid as a radiating atmosphere of a star such as in the
Vaidya metric.\cite{Vaidya53} Apparently, we can interpret it as to describe
a star immersed in a homogenous and isotropic background radiation.

The existence of the heat flow term in the energy-momentum tensor (\ref
{T-heat}) implies that there must be a temperature gradient in the field and
a heat interchanges between the star and the background radiation. Using (%
\ref{clos-to2}) and (\ref{clos-to3}) in (\ref{T-solu}) gives 
\begin{equation}
\sqrt{B(r)}T(t,r)-\sqrt{B_R}T_R(t)=\frac{\lambda M}{4\pi \kappa (1-\lambda t)%
}\left( \frac 1r-\frac 1R\right) \left[ 1+O\left( \varepsilon \right)
\right] \quad .  \label{T-appr}
\end{equation}
If we choose $R$ as the radius of the star, then in the exterior $r>R$,
equation (\ref{T-appr}) implies that if $\lambda >0$ then $\sqrt{B(r)}T(t,r)<%
\sqrt{B_R}T_R(t)$. Be aware that the thermal equilibrium corresponds to $%
\sqrt{B(r)}T(t,r)=\sqrt{B_R}T_R(t)$ for which there is no heat flow.\cite
{MTWbook} So we conclude that the heat flows outwards if $\lambda >0$ and
inwards if $\lambda <0$. This agree with the original relation (\ref{heat})
in which $q^1>0$ for $\lambda >0$ and $q^1<0$ for $\lambda <0$ since $%
B^{\prime }>0$ according to (\ref{clos-to2}). Meanwhile, from the metric (%
\ref{clos-to1}) and the result (\ref{rho-appr}) we also see that if $\lambda
>0$, then as the time $t$ increases, the 3D space contracts and the energy
density of the outside fluid increases. All these properties are physically
reasonable.

\section{DISCUSSION}

As we have concluded in section 4 that the 4D solution discussed in this
paper can be interpreted as an exterior solution of a spherical source such
as a star which has a non-zero temperature and is immersed in a homogenous
and isotropic background radiation. A natural candidate for this kind of
radiation is the cosmic background radiation, for which the temperature is
about $T_0\approx 2.7K$ at present days with an energy density around $\rho
_b\approx 4.0\times 10^{-13}erg$ $cm^{-3}$. Thus, by using (\ref{rho}), we
determine the constant $\lambda $ as 
\begin{equation}
\left| \lambda \right| ^{-1}\approx \sqrt{\frac{3c^4}{4\pi G\rho _b}}\approx
2.7\times 10^{30}cm\approx 1.8\times 10^{17}AU\;.  \label{lambda}
\end{equation}
So $\lambda $ is of the order of the Hubble constant. Now we wish to know if
the close-to-Schwarzschild solution given in (\ref{clos-to1})-(\ref{clos-to3}%
) can describe our solar system. That is, we need to calculate and compare
orders of the two terms $GM/r$ and $\lambda ^2r^2$ appeared in (\ref
{clos-to2}) and (\ref{clos-to3}). The average distance between the Sun and
the nearest planet Mercury is about $0.387$ AU. So we have 
\begin{equation}
\frac{GM_{\odot }}{R_{Merc}c^2}\approx 2.55\times 10^{-8}\;,\quad \left|
\lambda \right| R_{Merc}\approx 2\times 10^{-18}\quad .  \label{Mercury}
\end{equation}
The average distance between the Sun and the Pluto is about $39.53$ AU. So 
\begin{equation}
\frac{GM_{\odot }}{R_{Pluto}c^2}\approx 2.5\times 10^{-10}\;,\quad \left|
\lambda \right| R_{Pluto}\approx 2\times 10^{-16}\quad .  \label{Pluto}
\end{equation}
Therefore we conclude that in a wide range in our solar system, we have $%
\left| \lambda \right| r\ll GM/r$. So $\lambda ^2r^2$ is of the order higher
than the post-Newtonian order. This implies that the contributions of the
cosmic background radiation to all of the known solar system experiments\cite
{Will93} are negligible. However, there may have other ways to detect
possible new effects of the solution. For example, according to the
time-dependent metric (\ref{clos-to1}) and the value of $\lambda $ in (\ref
{lambda}), the radius of the central star should be contracting or
expanding, depending on whether $\lambda $ is positive or negative, with a
relative rate at present time being 
\begin{equation}
\left( \frac{\stackrel{.}{R}}R\right) _0=-\lambda c\approx -3.5\times
10^{-13}yer^{-1}\;,\qquad \text{for }\lambda >0\quad .
\end{equation}
It is worth to study whether this kind of effect is observable.

\section*{ACKNOWLEDGMENTS}

We thank Paul Wesson, Mark Roberts and Guowen Peng for comments. This work
was supported by NSF of P. R. China under grant 19975007.

\end{document}